\documentclass[10pt]{article}
\usepackage{amsmath,amsfonts,amssymb,cite}
\newcommand{\be}{\begin{equation}}
\newcommand{\ee}{\end{equation}}
\newcommand{\ba}{\begin{eqnarray}}
\newcommand{\ea}{\end{eqnarray}}

\newcommand{\la}[1]{\label{#1}}

\def\gl#1{(\ref{#1})}

\date{}
\begin{document}
\title{Polynomial Supersymmetry for Matrix Hamiltonians}
\author{A.V. Sokolov\footnote{E-mail: avs\_avs@rambler.ru.}\\ \\{\it Department of
Theoretical Physics, Saint-Petersburg State University,}\\{\it  Ulianovskaya ul., 1,  Saint-Petersburg
198504, Russia}}

\maketitle

\abstract{We study intertwining relations for matrix one-dimensional, in general, non-Hermitian Hamiltonians by matrix differential operators of arbitrary order. It is established that for any matrix intertwining operator $Q_N^-$ of minimal order $N$ there is a matrix operator $Q_{N'}^+$ of different, in general, order $N'$ that intertwines the same Hamiltonians as $Q_N^-$ in the opposite direction and such that the products $Q_{N'}^+Q_N^-$ and $Q_N^-Q_{N'}^+$ are identical polynomials of the corresponding Hamiltonians. The related polynomial algebra of supersymmetry is constructed. The problems of minimization and of reducibility of a matrix intertwining operator are considered and the criteria of minimizability and of reducibility are presented. It is shown that there are absolutely irreducible matrix intertwining operators, in contrast to the scalar case.}

\


\section{Introduction}

Supersymmetric matrix models in Quantum Mechanics arise for spectral design in description of motion of spin particles in external fields and of scattering of particles with strong coupling of channels. The simplest cases of Darboux transformations for matrix Hamiltonians and the corresponding supersymmetry algebras are considered in several papers, in particular, in \cite{acd88,acd90-1,aisv91,canio92,canio93,takahiro93,hgb95,lrsfv98,dc99,tr99,iknn06,fmn10,nk11}. Matrix $n\times n$ differential operators of the first order and matrix $2\times2$ differential operators of the second order which intertwine Hermitian Hamiltonians and the corresponding algebras of supersymmetry were studied in \cite{acni97}. In \cite{gove98} the formulae were proposed that allow us to construct for a given matrix $n\times n$, in general, non-Hermitian Hamiltonian a new matrix $n\times n$ Hamiltonian and a matrix $n\times n$ linear differential operator of the $N$-th order with the identity matrix coefficient at $({d\over{dx}})^N$ intertwining these Hamiltonians.

The formulae of \cite{gove98} are very general since they provide us with the opportunity to build any matrix intertwining operator with the identity matrix coefficient at $d\over{dx}$ in the highest degree, but at the same time the results of \cite{gove98} have some shortcomings. Firstly, the  formulae considered in \cite{gove98} are constructed in terms of a basis in a subspace which is invariant with respect to initial Hamiltonian, {\it i.e.} in terms of columns of $n\times nN$ matrix-valued solution ${\bf\Psi}(x)$ of the equation \begin{equation}H_+{\bf\Psi}={\bf\Psi}\Lambda,\la{gove1}\end{equation} where $H_+$ and $\Lambda$ are initial Hamiltonian and $nN\times nN$ constant matrix respectively. In the present work we demonstrate that in order to construct any matrix $n\times n$ intertwining operator of arbitrary order $N$ with arbitrary constant nondegenerate matrix coefficient at $({d\over{dx}})^N$ it is sufficient in fact to use only such matrices ${\bf\Psi}(x)$ that the matrix $\Lambda^t$ has a normal (Jordan) form. It is evident that in this case the columns of ${\bf\Psi}(x)$ are formal vector-eigenfunctions or formal associated vector-functions of a given initial Hamiltonian~$H_+$, where the word ``formal'' means that these vector-functions are not necessarily normalizable.

Secondly, the formulae of \cite{gove98} are rather complicated, in particular,  because of the use of {\it quasideterminants} introduced in \cite{gere}. More useful formulae for a matrix intertwining operator with the identity matrix coefficient at $d\over{dx}$ in the highest degree and for the potential of the related new Hamiltonian, in terms of usual determinants, are obtained in a very complex way in \cite{sampe03} for the
partial case where all columns of ${\bf\Psi}(x)$ are formal vector-eigenfunctions of $H_+$. We emphasize that the use of only formal vector-eigenfunctions of $H_+$ as columns in ${\bf\Psi}(x)$ leads to the fact that the set of intertwining operators, which can be constructed with the help of the formulae of \cite{sampe03}, is much narrower than the set of intertwining operators which can be received due to the formulae of \cite{gove98}.
In the present paper we derive  in a simple way the formulae for any matrix $n\times n$ intertwining operator of arbitrary order $N$ with an arbitrary constant nondegenerate matrix coefficient at $({d\over{dx}})^N$ and for the potential of related new Hamiltonian in terms of usual determinants. Our formulae in partial case of \cite{sampe03} correspond to the formulae of \cite{sampe03}.

The third shortcoming of \cite{gove98} which is present in \cite{sampe03} as well is the absence of any condition that guarantees (i) implementability of the described procedure of constructing of a matrix intertwining operator and of the corresponding new matrix Hamiltonian and (ii)~smoothness (the absence of pole(s)) for the matrix coefficients of constructed intertwining operator and for the potential of the corresponding new Hamiltonian. Moreover, Theorem~1 from \cite{gove98} containing a sufficient condition of existence of a matrix intertwining operator is wrong in view of the following arguments. This theorem asserts that for any $nN$-dimensional invariant with respect to matrix $n\times n$ initial Hamiltonian $H_+$ subspace $V$ of $n$-dimensional vector-functions there exists the intertwining matrix $n\times n$ differential operator $Q_N^-$ of the order $N$ and a final matrix $n\times n$ Hamiltonian $H_-$ such that $\ker Q_N^-= V$ and $Q_N^-H_+=H_-Q_N^-$.  Incorrectness of this statement takes place in view of the following simple counterexample.
Let us assume that $h_1=-({d\over{dx}})^2+v_1(x)$ and $h_2=-({d\over{dx}})^2+v_2(x)$ are  scalar, in general, non-Hermitian Hamiltonians and $\varphi_1(x)$ and $\varphi_2(x)$ are formal ({\it i.e.} not necessarily from $L_2(\Bbb R)$) eigenfunctions of $h_1$ for the spectral values $\lambda_1$ and $\lambda_2$ respectively such that
\[h_1\varphi_j=\lambda_j\varphi_j,\quad\lambda_j\in{\Bbb C},\quad j=1,2,\qquad \varphi'_1(x)\varphi_2(x)-\varphi_1(x)\varphi'_2(x)\not\equiv0.\] Then, the subspace with the basis \begin{equation}\Phi_j(x)=\begin{pmatrix}\varphi_j(x)\\0\end{pmatrix},\qquad j=1,2\la{bas1}\end{equation} is  obviously  invariant with respect to the matrix Hamiltonian \[H_+={\rm{diag}}\,(h_1,h_2),\] but it is not hard to see that there is no a matrix $2\times2$ linear differential operator of first order with the identity matrix coefficient or with any other nonzero coefficient at ${d\over{dx}}$, the kernel of which has the basis \gl{bas1}. In addition, for the solution \[{\bf\Psi}(x)=\begin{pmatrix}\varphi_1(x)&\varphi_2(x)\\0&0\end{pmatrix},\qquad\Lambda=\begin{pmatrix}\lambda_1&0\\0&\lambda_2\end{pmatrix}\] of the equation \gl{gove1} the procedures from \cite{gove98} and \cite{sampe03} of constructing of a matrix intertwining operator and the corresponding new Hamiltonian are not implementable.

One can check that the mentioned theorem from \cite{gove98} and its proof would be correct if in the formulation of this theorem to add to properties of $V$  the condition that the Wronskian of elements of a basis in $V$ is different from identical zero. It will be shown in our paper that the condition that the Wronskian of a set of formal vector-functions and formal associated vector-functions of $H_+$ does not vanish on the real axis together with the condition that the matrix $\Lambda^t$ corresponding to this set has a normal (Jordan) form guarantees (i) implementability of the procedure, described in our paper, of constructing of a matrix intertwining operator and of the corresponding new matrix Hamiltonian  and (ii) smoothness (absence of pole(s)) for the matrix coefficients of constructed intertwining operator and for the potential of the corresponding new Hamiltonian.

The fourth shortcoming of \cite{gove98} is the absence of a direct answer to the question about possibility to construct any matrix intertwining operator with the identity matrix coefficient at $d\over{dx}$ in the highest degree with the help of the procedure described there, though the positive answer to this question follows from the corrected version of the  theorem from \cite{gove98} mentioned above and from the fact that a matrix linear differential operator with the identity matrix coefficient at $d\over{dx}$ in the highest degree is uniquely defined by its kernel.

The paper \cite{pepusa11} is devoted to the generalization of results of \cite{sampe03} on the case with a degenerate matrix coefficient at $d\over{dx}$ in the highest degree in an intertwining operator. In \cite{tanaka11} it is proposed to consider a supersymmetry
 for two matrix $n\times n$, in general, non-Hermitian Hamiltonians of Schr\"odinger form $H_+$ and $H_-$ generated by two matrix $n\times n$ linear differential operators $Q_N^+$ and $Q_N^-$ of the same order $N$ with constant coefficients proportional to the identity matrix at $({d\over{dx}})^N$. It is supposed that the operators $Q_N^+$ and $Q_N^-$  intertwine the Hamiltonians $H_+$ and $H_-$ in the opposite directions so that the products $Q_N^+Q_N^-$ and $Q_N^-Q_N^+$ are the same polynomials with matrix coefficients of the Hamiltonians $H_+$ and $H_-$ respectively. Moreover, the intertwining operators $Q_N^+$ and $Q_N^-$ are assumed to be mutually conjugated by certain unnatural operation which is, in general, neither Hermitian conjugation nor transposition. Thus, intertwining of the Hamiltonians by one of these operators is not, in general, consequence of intertwining of the Hamiltonians by another of these operators, even in the case where both Hamiltonians are Hermitian conjugated or symmetric with respect to transposition. Hence, these intertwining operators lead, in general, to independent restrictions on the considered system. In \cite{tanaka11} it was not offered any method for finding of the potentials and the coefficients of considered Hamiltonians and intertwining operators respectively. Only in the case $n=N=2$ it was found a general solution of  intertwining relations under additional assumption that both potentials and all coefficients of intertwining operators are Hermitian.

One of main results of the present paper is that for any matrix $n\times n$ linear differential operator $Q_N^-$ of arbitrary order $N$ with nondegenerate matrix coefficient at $({d\over{dx}})^N$ that intertwines two matrix $n\times n$, in general,  non-Hermitian Hamiltonians $H_+$ and $H_-$ of Schr\"odinger form, there is a matrix $n\times n$ linear differential operator $Q_{N'}^+$ of the order $N'$ different, in general, from $N$ that intertwines the same Hamiltonians in the opposite direction and such that the product $Q_{N'}^+Q_N^-$ is a polynomial (coefficients of which are  numbers and not matrices as in \cite{tanaka11}) of $H_+$. Moreover, if there is no a nonzero intertwining operator of the order less than $N$ then the product $Q_N^-Q_{N'}^+$ is the same polynomial of $H_-$. Polynomial supersymmetry algebra constructed of the  Hamiltonians $H_+$ and $H_-$ and of the intertwining operators $Q_N^-$ and $Q_{N'}^+$ is presented in this paper as well.

All papers  mentioned above are devoted in fact (in regard to Darboux transformations) to the case with one spatial variable, only \cite{aisv91} briefly touches also the cases with two and three spatial variables. The latter cases are considered in \cite{ione03} too.

It is evident that a product of an intertwining operator and a polynomial of the corresponding Hamiltonian with the coefficients being symmetry operators for this Hamiltonian is again an intertwining operator for the same pair of Hamiltonians. Thus, there is the problem of minimization of an intertwining operator by means of removing from it a superfluous factor polynomial in the corresponding Hamiltonian. The criterion is presented in this paper of a weak minimizability of a matrix differential intertwining operator, {\it i.e.} of the possibility to separate from this operator a nonconstant polynomial of the corresponding Hamiltonian coefficients of which are a numbers (not matrices). This criterion is analogous to one of \cite{anso03,ancaso07}.

The problem of reducibility of a differential intertwining operator, {\it i.e.} of possibility to represent it in the form of a product of differential intertwining operators of lower orders with smooth coefficients\footnote{If the intertwined Hamiltonians are either Hermitian or symmetric with respect to transposition or all elements of their potentials are real-valued then one can impose the respective restrictions on the potentials of all intermediate Hamiltonians and a suitable additional restrictions on the coefficients of all intertwining operators of lower orders.}, is important for the theory of intertwining operators and the theory of spectral design (see, for example, \cite{acdi95,samsonov99,anca04,anso06,sokolov07,sokolov10} and references therein) since reducibility of an intertwining operator allows us to reduce a complicated transformation of an initial Hamiltonian to a final Hamiltonian described by this intertwining operator to a chain of more simple transformations described by an intertwining operators of lower orders. We  present in this paper a criterion of reducibility for a matrix differential intertwining operator. In addition, it is shown that in contrast to the scalar case $n=1$ (see, for example, Lemma 1 in \cite{anso03}) in the matrix case there are absolutely irreducible matrix differential intertwining operators, {\it i.e.} intertwining operators which cannot be represented in the form of a product of intertwining operators of lower orders even with a pole singularity(-ies) into coefficients.

This paper is based on the report \cite{sokolov12} and its goal is to present briefly and mostly without proofs our new results on intertwining of matrix Hamiltonians of Schr\"odinger form by matrix linear differential operators and on polynomial supersymmetry algebra  constructed of such operators and Hamiltonians.

The paper is organized as follows. The basic definitions and notation are presented in Section 2. Section~3 is devoted to description of constructing of any matrix $n\times n$ differential intertwining operator of the $N$-th order with arbitrary constant nondegenerate matrix coefficient at $\big({d\over{dx}}\big)^N$ and of the corresponding final Hamiltonian in terms of formal vector-eigenfunctions and formal associated vector-functions of initial matrix $n\times n$ Hamiltonian. Section 4 includes the definition of minimizability of a matrix differential intertwining operator (in the sense mentioned above as a weak minimizability) and the criterion of minimizability for such operator.  Section 5 contains the results on existence for any matrix intertwining operator $Q_N^-$ of the order $N$ a ``conjugate'' matrix intertwining operator $Q_{N'}^+$ and on polynomial algebra of supersymmetry with these operators. The definitions of reducible, irreducible and absolutely irreducible matrix differential intertwining operators, the criterion of reducibility for such operators and an example of absolutely irreducible matrix $2\times 2$ differential intertwining operator of the $2$-nd order (generalization of this example for arbitrary $n\geqslant2$ and $N\geqslant2$ is straightforward) are given in Section 6. Conclusions contain the list of problems which can be investigated in the future papers.

\section{Basic definitions and notation}

Let us consider two defined on the entire axis matrix $n\times n$ Hamiltonians
\[H_+=-I_n\partial^2+V_+(x),\quad H_-=-I_n\partial^2+V_-(x),\qquad
\partial\equiv{d/{dx}},\la{h+h-2.1}\] where $I_n$ is the identity matrix and $V_+(x)$ and $V_-(x)$ are square matrices, all elements of which are sufficiently smooth and, in general, complex-valued functions. We suppose that these Hamiltonians are
{\it intertwined} by a matrix linear differential operator $Q_N^-$, so that
\begin{equation} Q_N^-H_+=H_-Q^-_N,\qquad
Q_N^-=\sum\nolimits_{j=0}^NX^-_j(x)\partial^j,\la{splet}\end{equation}
where $X^-_j(x)$, $j=0$, \dots, $N$ are as well square matrices of
$n$-th order, all elements of which are sufficiently smooth and, in
general, complex-valued functions.

Transforming the left- and right-hand sides of \gl{splet} into
expansions in powers of $\partial$ and equating coefficients of
equal powers of $\partial$ in these sides, we obtain the following
equations for the cases with the coefficients for $\partial^{N+1}$
and $\partial^N$:
\[-X^-_{N-1}(x)=-2X^{-\,\prime}_{N}(x)-X^-_{N-1}(x),\]
\[-X^-_{N-2}(x)+X^-_N(x)V_+(x)=-X^{-\,\prime\prime}_N(x)
-2X^{-\,\prime}_{N-1}(x)-X^-_{N-2}(x)+V_-(x)X^-_N(x).\] It follows
from the first of these equations that $X^-_N(x)$ is a constant
matrix and, thus, the second of these equations takes the form
\begin{equation}X^-_NV_+(x)=-2X^{-\,\prime}_{N-1}(x)+V_-(x)X^-_N.\la{v12}\end{equation}
In the what follows we restrict ourselves by the case $\det X_N^-\ne0$. In this case one can find from \gl{v12} the
matrix potential $V_-(x)$ in terms of $V_+(x)$ and $X^-_{N-1}(x)$,
\begin{equation}V_-(x)=X^-_NV_+(x)(X^-_N)^{-1}+2X^{-\,\prime}_{N-1}(x)(X^-_N)^{-1}.\la{vmp2.1}\end{equation}

Existence of a ``conjugate'' matrix $n\times n$ intertwining operator $Q_M^+$ for a given matrix intertwining operator $Q_N^-$ such that \begin{equation} H_+Q_M^+=Q^+_MH_-,\qquad
Q_M^+=\sum\nolimits_{j=0}^MX^+_j(x)\partial^j,\la{q+int}\end{equation}
is not evident in general but is evident in the following cases:
\renewcommand{\labelenumi}{(\theenumi)}
\begin{enumerate}
\item $H_+^\dag\!=\!H_+,\,\,H^\dag_-\!=\!H_-\quad\Rightarrow\quad H_+Q_N^+\!=\!Q_N^+H_-,\quad
Q_N^+\!=\!Q_N^{-\,\dag}\!=\!\sum_{j=0}^N(-\partial)^jX_j^{-\,\dag}(x),$\hfill\\ where $\dag$ denotes Hermitian conjugation;
\item $H_+^t\!=\!H_+,\,\,H^t_-\!=\!H_-\quad\Rightarrow\quad H_+Q_N^+\!=\!Q_N^+H_-,\quad
Q_N^+\!=\!Q_N^{-\,t}\!=\!\sum_{j=0}^N(-\partial)^jX_j^{-\,t}(x),$\hfill\\ where $t$ denotes transposition;
\item \qquad$H_+^\ast=H_-\quad\Rightarrow\quad H_+Q_N^+=Q_N^+H_-,\qquad Q_N^+=Q_N^{-\,\ast}=\sum_{j=0}^NX_j^{-\,\ast}(x)\partial^j,$\\ where $*$ denotes complex conjugation.
\end{enumerate}
Existence of a ``conjugate'' matrix intertwining operator of the type \gl{q+int} for any matrix intertwining operator~$Q_N^-$ is guaranteed by the results of Section \ref{ecio}.

By virtue of the intertwining \gl{splet} the kernel of $Q_N^-$ is an invariant
subspace for~$H_+$: \[H_+\ker Q_N^-\subset\ker
Q_N^-.\] Hence, for any basis $\Phi^-_1(x)$, \dots, $\Phi^-_d(x)$ in $\ker Q_N^-$, $d=\dim\ker Q_N^-=nN$ there is a constant
square matrix $T^+\equiv\|T^+_{ij}\|$ of the $d$-th order such that
\[H_+\Phi^-_i=\sum\nolimits_{j=1}^dT^+_{ij}\Phi^-_j,\qquad i=1,\ldots,d.
\la{tm}\]

A basis in the kernel of an intertwining operator $Q_N^-$ in which the
matrix $T^+$ has a normal (Jordan) form is called a {\it
canonical basis}. Elements of a canonical basis are called a {\it
transformation vector-functions}.

If a Jordan form of the matrix $T^+$ has block(s) of order higher than one, then the corresponding canonical basis contains not only formal vector-eigenfunctions of $H_+$ but also formal associated vector-function(s) of $H_+$ which are defined as follows (see \cite{naim}).

A finite or infinite set of vector-functions $\Phi_{m,i}(x)$, $i=0$, 1, 2, \dots\, is called a {\it chain of formal associated
vector-functions} of $H_+$ for a spectral value $\lambda_m$ if \[H_+\Phi_{m,0}\!=\!\lambda_m\Phi_{m,0},
\quad\Phi_{m,0}(x)\!\not\equiv\!0,
\qquad (H_+\!-\!\lambda_mI_n)\Phi_{m,i}\!=\!\Phi_{m,i-1},\quad i\!=\!1,2,3,\ldots\,.\] The vector-function $\Phi_{m,i}(x)$ in this case
is called {\it a formal associated vector-function of $i$-th order} of the Hamiltonian $H_+$ for the spectral value $\lambda_m$, $i=0$, 1, 2, \dots\, and $\Phi_{m,0}(x)$ is called as well a formal vector-eigenfunction of $H_+$ for the same spectral value. The term ``formal'' emphasizes that the vector-function $\Phi_{m,i}(x)$ is not necessarily normalizable, $i=0$, 1, 2, \dots\,.

\section{Constructing of a matrix intertwining operator in terms of transformation vector-functions \la{ioNgc}}

Let us consider a set of formal associated vector-functions \[\Phi_l^-(x)\equiv\big(
\varphi^-_{l1}(x),\varphi^-_{l2}(x),\ldots,\varphi^-_{ln}(x)\big)^t,\qquad l=1,\ldots,nN,\qquad n,N\in{\Bbb N}\la{phl5}\]
of a matrix $n\times n$ Hamiltonian $H_+$ such that this set can be divided into a chains of formal associated vector-functions of $H_+$ for different, in general, spectral values of $H_+$ and the Wronskian $W(x)$ of all $\Phi^-_l(x)$, $l=1$, \dots, $nN$ does not vanish on the real axis. There is the only matrix $n\times n$ linear differential operator $Q_N^-$ of the $N$-th order with arbitrarily fixed constant nondegenerate matrix coefficient at $\partial^N$ such that $\ker Q_N^-$ contains all vector-functions $\Phi_l^-(x)$, $l=1$, \dots, $nN$. This operator can be found with the help of the following evident explicit formula,
\[Q_N^-={1\over{W(x)}}\,
X_N^-\begin{vmatrix}\varphi^-_{11}&\dots&\varphi^-_{1n}\quad \varphi^{-\prime}_{11}&\dots&\varphi^{-\prime}_{1n}&\ldots&(\varphi^-_{11})^{(N-1)}&\dots&(\varphi^-_{1n})^{(N-1)}&(\Phi^-_1)^{(N)}\\
\varphi^-_{21}&\dots&\varphi^-_{2n}\quad
\varphi^{-\prime}_{21}&\dots&\varphi^{-\prime}_{2n}&\ldots&
(\varphi^-_{21})^{(N-1)}&\dots&(\varphi^-_{2n})^{(N-1)}&(\Phi^-_2)^{(N)}\\
\vdots&\ddots&\vdots\qquad\vdots&\ddots&\vdots&\ddots&\vdots&\ddots&\vdots&\vdots\\
\varphi^-_{nN,1}\!\!\!\!&\dots&\!\!\!\!\varphi^-_{nN,n}\,\,\varphi^{-\prime}_{nN,1}\!\!\!\!&\dots&\!\!\!\!\varphi^{-\prime}_{nN,n}\!\!\!\!&\ldots&\!\!\!\!(\varphi^-_{nN,1})^{(N-1)}\!\!\!\!&\dots&\!\!\!\!(\varphi^-_{nN,n})^{(N-1)}\!\!&\!\!(\Phi^-_{nN})^{(N)}\\
P_1&\ldots&P_n\quad P_1\partial&\ldots&P_n\partial&\ldots&P_1\partial^{N-1}&\ldots&P_n\partial^{N-1}&I_n\partial^N
\end{vmatrix}\!,\] \begin{equation}P_l\Phi=\varphi_l,\qquad\forall\,\,\Phi(x)\equiv\big(\varphi_1(x),\varphi_2(x),\ldots,\varphi_n(x)\big)^t,\qquad l=1,\ldots,n,\la{qNrep5}\end{equation}
where during calculation of the determinant in each of its terms the corresponding of the operators $P_1$, \dots, $P_n$, $P_1\partial$, \dots, $P_n\partial$, $P_1\partial^{N-1}$, \dots, $P_n\partial^{N-1}$, $I_n\partial^N$ must be placed on the last position.
It is not hard to see in view of \gl{qNrep5} that $l$-th column of the matrix coefficient $X_j^-(x)$ of $Q_N^-$ (see \gl{splet}) is equal to
\[-{1\over{W(x)}}\, X_N^-\,\,
\begin{matrix}\big|&\!\!\!\varphi^-_{11}&\dots&\varphi^-_{1n}&\varphi^{-\prime}_{11}&\dots&
\varphi^{-\prime}_{1n}&\ldots\\
\Big|&\!\!\!\varphi^-_{21}&\dots&\varphi^-_{2n}&\varphi^{-\prime}_{21}&\dots&\varphi^{-\prime}_{2n}&\ldots\\
\Bigg|&\!\!\!\vdots&\ddots&\vdots&\vdots&\ddots&\vdots&\ddots\\
\Big|&\!\!\!\varphi^-_{nN,1}&\dots&\varphi^-_{nN,n}&\varphi^{-\prime}_{nN,1}&\dots&\varphi^{-\prime}_{nN,n}&\ldots
\end{matrix}\qquad\qquad\qquad\qquad\qquad\qquad\quad\]

\[\quad\qquad \begin{matrix}(\varphi^-_{1,l-1})^{(j)}&(\Phi^-_1)^{(N)}&
(\varphi^-_{1,l+1})^{(j)}&\ldots&(\varphi^-_{11})^{(N-1)}&\dots&(\varphi^-_{1n})^{(N-1)}&\Big|\\
(\varphi^-_{2,l-1})^{(j)}&(\Phi^-_2)^{(N)}&(\varphi^-_{2,l+1})^{(j)}&\ldots&
(\varphi^-_{21})^{(N-1)}&\dots&(\varphi^-_{2n})^{(N-1)}&\Big|\\
\vdots&\vdots&\vdots&\ddots&\vdots&\ddots&\vdots&\Bigg|\\
(\varphi^-_{nN,l-1})^{(j)}&(\Phi^-_{nN})^{(N)}&(\varphi^-_{nN,l+1})^{(j)}&\ldots&(\varphi^-_{nN,1})^{(N-1)}\!&\dots&(\varphi^-_{nN,n})^{(N-1)}\!\!\!&\Big|
\end{matrix}\]

\begin{equation} l=1,\ldots,n,\qquad j=0,\dots,N-1.\la{tilstx5}\end{equation}
We emphasize that the condition that the Wronskian $W(x)$ does not vanish on real axis provides in view of \gl{qNrep5} and \gl{tilstx5} existence for $Q_N^-$ and smoothness (absence of pole(s)) for the matrix-valued functions $X_0^-(x)$, \dots, $X_{N-1}^-(x)$.

Existence of a matrix $n\times n$ Hamiltonian $H_-$ of Schr\"odinger form which is intertwined with $H_+$ by $Q_N^-$ in accordance with \gl{splet} can be proved in the same way as in the proof of Theorem 1 in \cite{gove98}. The potential $V_-(x)$ of the Hamiltonian $H_-$ can be found with the help of \gl{tilstx5} for $j=N-1$ and the relation \gl{vmp2.1}. Thus, the potential $V_-(x)$ is smooth as well.
The partial cases of the representation of $Q_N^-\Phi$ for arbitrary $n$-dimensional vector-function $\Phi(x)$ with the help of \gl{qNrep5} and of the representation of $V_-(x)$ with the help of \gl{tilstx5} for $j=N-1$ and \gl{vmp2.1} when $X_N^-=I_n$ and all vector-functions $\Phi_l^-(x)$, $l=1$, \dots, $nN$ are formal vector-eigenfunctions of the Hamiltonian $H_+$ are contained in \cite{sampe03}.

The fact that for any matrix $n\times n$ initial Hamiltonian $H_+$ of Schr\"odinger form any matrix $n\times n$ linear differential intertwining operator of arbitrary order $N$ with arbitrary constant nondegenerate matrix coefficient at $\partial^N$ can be constructed with the help of the procedure described in this section follows from the facts that (i) for any such operator there is a canonical basis in its kernel, Wronskian of which does not vanish on the real axis and (ii) a matrix $n\times n$ linear differential operator of $N$-th order with a given constant nondegenerate matrix coefficient at $\partial^N$ is uniquely determined by a basis in its kernel.

\section{Minimizability of a matrix intertwining operator}

It is evident that if to multiply $Q_N^-$ by a polynomial of the Hamiltonian,
\[Q_N^-\Big[\sum\nolimits_{l=0}^La_lH_+^l\Big]\equiv\Big[\sum\nolimits_{l=0}^{L}a_lH_-^l\Big]Q_N^-,\qquad a_l\in{\Bbb C},\quad l=0,\ldots,L,\]  then such product is again an intertwining operator for the same Hamiltonians:
\[\Big\{Q_N^-\Big[\sum\nolimits_{l=0}^La_lH_+^l\Big]\Big\}H_+=Q_N^-H_+\Big[\sum\nolimits_{l=0}^La_lH_+^l\Big]=H_-\Big\{Q_N^-\Big[\sum\nolimits_{l=0}^La_lH_+^l\Big]\Big\}.\]
Thus, the question arises about possibility to simplify an intertwining operator by separation from it a superfluous polynomial in the corresponding Hamiltonian factor. 

Let us present definition of minimizable and non-minimizable matrix intertwining operators.

\vskip1pc

\noindent{\bf Definition 1.} An intertwining operator $Q_N^-$ is called {\it minimizable} if this operator can be represented in the form
\[Q_N^-\!=\!P_M^-\Big[\sum\nolimits_{l=0}^La_lH_+^l\Big],\qquad a_l\!\in\!\Bbb C,\quad l\!=\!1, \ldots,L,\quad a_L\!\ne\!0,\quad1\!\leqslant\! L\!\leqslant\! N/2,\]
where $P_M^-$ is a matrix $n\times n$ linear differential operator of the $M$-th order, $M=N-2L$ that intertwines the Hamiltonians $H_+$ and $H_-$, so that $P_M^-H_+=H_-P_M^-$. Otherwise, the operator $Q_N^-$ is called {\it non-minimizable}.

\vskip1pc

The following criterion of minimizability takes place. A~matrix $n\times n$ nonzero intertwining operator $Q_N^-$ can be represented in the form
\[Q_N^-=P_M^-\prod_{l=1}^s(\lambda_lI_n-H_+)^{ k_l},\qquad\lambda_l\in{\Bbb C},\,\, k_l\in\Bbb N,\,\,
l=1, \ldots, s,\quad\lambda_l\ne \lambda_{l'}\Leftrightarrow l\ne l',\la{minim8}\]
where $P_M^-$ is a non-minimizable matrix $n\times n$ linear differential operator of the $M$-th order, $M=N-2\sum_{l=1}^sk
_l$ that intertwines the Hamiltonians $H_+$ and $H_-$, so that $P_M^-H_+=H_-P_M^-$,

\vskip0.5pc

\noindent if and only if
\renewcommand{\labelenumi}{\rm{(\theenumi)}}
\begin{enumerate}
\item all numbers $\lambda_l$, $l=1$, \dots, $s$ belong to the spectrum of the matrix $T^+$
and there are no equal numbers between them$;$
\item there are $2n$ Jordan blocks in a normal $($Jordan$)$ form of the matrix $T^+$ for any eigenvalue from the set $\lambda_l$, $l=1$, \dots, $s;$
\item there are no $2n$ Jordan blocks in a normal $($Jordan$)$ form of $T^+$ for any eigenvalue of this matrix that does not belong to the set $\lambda_l$, $l=1$, \dots, $s;$
\item $ k_l$ is the minimal of the orders of Jordan blocks corresponding to the eigenvalue $\lambda_l$ in a normal $($Jordan$)$ form of the matrix $T^+$, $l=1$, \dots, $s.$
\end{enumerate}

\section{On existence a ``conjugate'' intertwining operator and polynomial SUSY\la{ecio}}

Suppose that
\renewcommand{\labelenumi}{\rm{(\theenumi)}}
\begin{enumerate}
\item $\lambda_l$, $l=1$, \dots, $L$ is the set of all different eigenvalues of $T^+;$
\item $g_l^-$ is the geometric multiplicity of $\lambda_l$ in the spectrum of $T^+$, $l=1$, \dots, $L;$
\item $k_{l,j}^-$, $j=1$, \dots, $g_l^-$ are the orders of Jordan blocks corresponding to $\lambda_l$ in a Jordan form of $T^+$, $l=1$, \dots, $L;$
\item $\varkappa_l=\max_{1\leqslant j\leqslant g_l^-}k_{l,j}^-$, $l=1$,\dots, $L.$
\end{enumerate}
Then there is a non-minimizable linear differential operator $Q_{N'}^+$ of the order $N'=2(\varkappa_1+\ldots+\varkappa_L)-N$ with smooth coefficients that intertwines $H_+$ and $H_-$ as follows,
\[H_+Q_{N'}^+=Q_{N'}^+H_-\la{int'10}\]
and such that$:$
\[Q_{N'}^+Q_N^-=\prod\nolimits_{l=1}^L(H_+-\lambda_lI_n)^{\varkappa_l}.\la{pol10}\]
Moreover, if there is no a nonzero matrix linear differential operator $P_M^-$ of the order $M$, $M<N$ such that the following intertwining holds, \[P_M^-H_+=H_-P_M^-,\la{splet10.6}\]
then
\[Q_{N'}^+Q_N^-\!=\!{\cal P}_{(N+N')/2}(H_+),\quad Q_{N}^-Q_{N'}^+\!=\!{\cal P}_{(N+N')/2}(H_-),\quad{\cal P}_{(N+N')/2}(\lambda)\!\equiv\!\prod\nolimits_{l=1}^L(\lambda\!-\!\lambda_l)^{\varkappa_l}.\]

In the considered case with the help of the super-Hamiltonian \[{\bf H}=\begin{pmatrix}H_+&0\\0&H_-\end{pmatrix}\] and the nilpotent supercharges \[{\bf Q}=\begin{pmatrix}0&Q_{N'}^+\\0&0\end{pmatrix},\quad{\bf \bar Q}=\begin{pmatrix}0&0\\Q_N^-&0\end{pmatrix},
\qquad{\bf Q}^2={\bf \bar Q}^2=0\] one can construct the following polynomial algebra of supersymmetry:
\[\{{\bf Q},{\bf \bar Q}\}={\cal P}_{(N+N')/2}({\bf H}),\qquad[{\bf H},{\bf Q}]=[{\bf H},{\bf \bar Q}]=0.\la{supalg10}\]

\section{(Ir)reducibility of a matrix intertwining operator}

Let us present definitions of reducible, irreducible and absolutely irreducible matrix intertwining operators.

\vskip1pc

\noindent{\bf Definition 2.} An intertwining operator $Q_N^-$ is called {\it reducible} if there are a matrix $n\times n$ linear differential operators $K_{N-M}^-$ and~$P_M^-$ of the orders $N-M$ and $M$, $0<M<N$ respectively with smooth coefficients and a matrix $n\times n$ intermediate Hamiltonian of Schr\"odinger form $H_M$ with smooth potential such that the following relations hold,
\begin{equation}Q_N^-=K_{N-M}^-P_M^-,\qquad P_M^-H_+=H_MP_M^-,\qquad K_{N-M}^-H_M=H_-K_{N-M}^-.\la{interm7}\end{equation}
Otherwise the operator $Q_N^-$ is called {\it irreducible}.

\vskip1pc

\noindent{\bf Definition 3.} An intertwining operator $Q_N^-$ is called {\it absolutely irreducible} if for any $M$, $0<M<N$ there are no a matrix $n\times n$ linear differential intertwining operators $K_{N-M}^-$ and $P_M^-$ of the orders $N-M$ and $M$ respectively and a matrix $n\times n$ intermediate Hamiltonian of Schr\"odinger form $H_M$, even with the potential of $H_M$ and the coefficients of $K_{N-M}^-$ and $P_M^-$ possessing by a pole singularity(-ies), such that \gl{interm7} hold.

\vskip1pc

The following criterion of reducibility of a matrix intertwining operator takes place. A matrix $n\times n$ nonzero intertwining operator $Q_N^-$ is reducible if and only if there are a natural number $M<N$ and a vector-functions $\Phi_l^-(x)$, $l=1$, \dots, $nM$, $1\leqslant M<N$ belonging to $\ker Q_N^-$ such that these vector-functions can be divided into a chains of formal associated vector-functions of $H_+$ and their Wronskian does not vanish on the real axis.

\vskip1pc

In contrast to the scalar case $n=1$ where absolutely irreducible intertwining operators absent (see, for example, Lemma 1 in \cite{anso03}) there are in the matrix case with any $n\geqslant2$ absolutely irreducible intertwining operators of any order.
Restrict ourselves to the case $n=N=2$ and consider two chains of associated functions of two scalar Hamiltonians $h_1$ and $h_2$ for the same spectral value $\lambda_0\in\Bbb C$:
\begin{eqnarray}h_1\varphi_{1,0}=\lambda_0\varphi_{1,0},&\qquad&(h_1-\lambda_0)\varphi_{1,l}=\varphi_{1,l-1},\quad l=1,2,3,\nonumber\\
h_2\varphi_{2,0}=\lambda_0\varphi_{2,0},&\qquad&(h_2-\lambda_0)\varphi_{2,1}=\varphi_{2,0}\nonumber\end{eqnarray}
such that the Wronskians \[W_{1}(x)\equiv\varphi_{1,0}\varphi'_{1,1}-\varphi'_{1,0}\varphi_{1,1},\qquad
W_{2}(x)\equiv\varphi_{2,0}\varphi'_{2,1}-\varphi'_{2,0}\varphi_{2,1}\] do not vanish on the real axis. Then the vector-functions
\[\Phi_0^-=\begin{pmatrix}\varphi_{1,0}\\0\end{pmatrix},\quad\Phi_1^-=\begin{pmatrix}\varphi_{1,1}\\0\end{pmatrix},\quad\Phi_2^-=\begin{pmatrix}\varphi_{1,2}\\\varphi_{2,0}\end{pmatrix},\quad\Phi_3^-=\begin{pmatrix}\varphi_{1,3}\\\varphi_{2,1}\end{pmatrix},\] form a chain of formal associated vector-functions of the matrix Hamiltonian \[H_+={\rm{diag}}\,(h_1,h_2)\] for the spectral value $\lambda_0$,
\[H_+\Phi_0^-=\lambda_0\Phi_0^-,\qquad (H_+-\lambda_0I_2)\Phi_l^-=\Phi_{l-1}^-,\qquad l=1,2,3\]  and the following equalities for the Wronskian of  $\Phi_0^-(x)$, \dots, $\Phi_3^-(x)$ hold,
\[{\large\begin{vmatrix}\varphi_{1,0}&0&\varphi'_{1,0}&0\\
\varphi_{1,1}&0&\varphi'_{1,1}&0\\
\varphi_{1,2}&\varphi_{2,0}&\varphi'_{1,2}&\varphi'_{2,0}\\
\varphi_{1,3}&\varphi_{2,1}&\varphi'_{1,3}&\varphi'_{2,1}
\end{vmatrix}}=-{\large\begin{vmatrix}\varphi_{1,0}&\varphi'_{1,0}&0&0\\
\varphi_{1,1}&\varphi'_{1,1}&0&0\\
\varphi_{1,2}&\varphi'_{1,2}&\varphi_{20}&\varphi'_{2,0}\\
\varphi_{1,3}&\varphi'_{1,3}&\varphi_{2,1}&\varphi'_{2,1}
\end{vmatrix}}=-W_{1}(x)W_{2}(x).\]
Thus, the Wronskian of $\Phi_0^-(x)$, \dots, $\Phi_3^-(x)$ does not vanish on the real axis and there exist (see Section~\ref{ioNgc}) the matrix $2\times2$ Hamiltonian of Schr\"odinger form $H_-$ and the matrix $2\times 2$ linear differential operator of the second order $Q_2^-$ intertwining $H_+$ and $H_-$ such that $\Phi_0^-(x)$, \dots, $\Phi_3^-(x)$ form a canonical basis in $\ker Q_2^-$. Absolute irreducibility of $Q_2^-$ takes place in view of the facts that any canonical basis in the kernel of possible separated from the right-hand side of $Q_2^-$ intertwining operator can be constructed of $\Phi_0^-(x)$ and $\Phi_1^-(x)$ but their Wronskian
\[\begin{vmatrix}\varphi_{1,0}&0\\\varphi_{1,1}&0\end{vmatrix}\equiv0.\]
Generalization of this construction for arbitrary $n$ and $N$ is straightforward.

\section*{Conclusions}

Let us present in conclusion the following list of questions and problems which could be investigated in future papers.
\renewcommand{\labelenumi}{\rm{(\theenumi)}}
\begin{enumerate}
\item By analogy with \cite{anso03} to introduce the notion of (in)dependence for matrix differential intertwining operators, to find a criterion of dependence for such operators and to investigate the questions on maximal number of independent matrix differential intertwining operators and on a basis of these operators.
\item To investigate the question on existence in the matrix case a matrix differential symmetry operator with properties analogous to the properties of an antisymmetric with respect to transposition non-minimizable symmetry operator in the scalar case \cite{anso03,anso09}.
\item To investigate in details different partial cases of intertwining relations, in particular, the case where both intertwined Hamiltonians are Hermitian, the case where both intertwined Hamiltonians are symmetric with respect to transposition and the case where all elements of the potentials of both intertwined Hamiltonians are real-valued.
\item By analogy with \cite{acdi95,anca04,samsonov99,anso06,sokolov10,ferneni00,tr89,dun98,khsu99,fermura03,fermiros02',fersahe03,samsonov06} to investigate and classify irreducible and, in particular, absolutely irreducible matrix differential intertwining operators.
\item To generalize the results to the case with degenerate matrix coefficient at $\partial$ in the highest degree in a matrix differential intertwining operator.
\end{enumerate}

\section*{Acknowledgments}

The author is grateful to A.A. Andrianov for critical reading of this paper and valuable comments. This work was supported by the SPbSU project 11.0.64.2010.

\end{document}